\documentclass[5p,times,number,sort&compress]{elsarticle}

\usepackage{amsmath,amssymb}
\usepackage[hyphens]{url}

\journal{Physics Letters B}

\begin{document}

\begin{frontmatter}

\title{$Z_{3}$ symmetry of the CKM and PMNS matrices}

\author[first]{Piotr Kielanowski}
\ead{piotr.kielanowski@cinvestav.mx}
\affiliation[first]{organization={Departamento de Física, Centro de
    Investigación y Estudios Avanzados},
  addressline={Av. Instituto Politécnico Nacional 2508},
  city={Ciudad de México},
  postcode={07000},
  country={Mexico}}

\author[second]{S. Rebeca Juárez Wysozka}
\ead{rebecajw@gmail.com}
\affiliation[second]{organization={Departamento de Física, Escuela
    Superior de Física y Matemáticas, Instituto Politécnico
    Nacional. U.P Adolfo López Mateos}, city={Ciudad de México},
  postcode={07739}, country={Mexico}}

\author[third]{Liliana Vázquez Mercado}
\ead{liliana.vmercado@academicos.udg.mx}
\affiliation[third]{organization={Departamento de Física, Centro
    Universitario de Ciencias Exactas e Ingenierías, Universidad de
    Guadalajara}, addressline={Av. Revolución 1500, Colonia Olímpica},
  city={Guadalajara}, postcode={44430}, state={Jalisco},
  country={Mexico}}

\begin{abstract}
  We develop for the CKM and PMNS matrices a new representation with
  special properties. It is obtained by splitting each of these
  matrices into two rotations by the angle ${\sim}2\pi/3$ and a
  universal diagonal matrix with elements, which are cubic roots
  of~1. Such a representation of the CKM and PMNS matrices may
  indicate the $Z_{3}$ symmetry to be present in the Yukawa sector of
  the~SM. Identical mathematical structure of the CKM and PMNS
  matrices is also an extension of the quark-lepton universality. In
  this approach the CP violation is a natural consequence of the
  structure of the Yukawa couplings. The CP violating phase is not a
  fitted parameter and its value is governed by the parameters of two
  rotations. The parameters of the diagonalizing matrices of the
  bi-unitary transformation do not exhibit a hierarchy, which means
  that the origins of the hierarchy of quark masses and of the CKM
  matrix elements are not the same.
\end{abstract}


\end{frontmatter}

\section{Introduction}
\label{sec:introduction}

In classical physics the invariance of all phenomena under the
discrete symmetries $P$, $T$ and $C$ was beyond doubt, so it was
considered certain that it should also be extended to quantum physics
at any level. This belief was unexpectedly broken by experimental
confirmation~\citep{PhysRev.105.1413} of the prediction of space parity
violation~\citep{PhysRev.104.254} in weak interactions. Almost
simultaneously the violation of the charge conjugation invariance $C$
was observed in the experiment measuring the polarization of muons in
the decays of charged pions~\citep{PhysRev.105.1415}. Next the
hypothesis of the combined CP invariance was experimentally proven to
be broken in neutral $K$ meson decays~\citep{PhysRevLett.13.138}. The
violation of the time reversal invariance~$T$ was observed for the
first time in the neutral~K system~\citep{199843}.

The discussion of the violation of discrete symmetries invariance was
gradually evolving in the theory of weak interaction as new
experimental results were emerging. The first theoretical attempt of
the description of $\beta$-decay was formulated by
Fermi~\citep{Fermi1934,10.1119/1.1974382} with only vector charged
currents interactions and it did not contain the breaking of discrete
symmetries. The parity violation through $V-A$ weak current was
introduced into the Fermi's theoretical description of weak
interactions by Sudarshan and Marshak in Refs.~\citep{Sudarshan,
  PhysRev.109.1860.2} and Feynman and Gell-Mann in
Ref.~\citep{PhysRev.109.193}. The next step in the theoretical
development was the reconciliation of universality of weak
interactions of weak decays of strange particles with $\beta$ decays
of muons by expressing the weak hadronic current as a mixture
described by one angle~$\theta$ of strangeness conserving current and
strangeness changing current~\citep{PhysRevLett.10.531}. CP~violation
can only be consistently described in the theory with three
generations of quarks and it was done in the framework of the Standard
Model (SM) by Kobayashi and Maskawa by introducing the $3\times3$
unitary mixing matrix for quark interactions with charged vector
bosons~\citep{10.1143/PTP.49.652}. We thus see that description of
charged current weak interactions by the Cabibbo-Kobayashi-Maskawa
(CKM) matrix is the result of the 40~years study of the properties of
elementary particles, with emphasis on the lepton-hadron universality.

For massive neutrinos in the lepton sector the mixing matrix, called
the Pontecorvo–Maki–Nakagawa–Sakata (PMNS)
matrix~\citep{Pontecorvo:1957cp, Pontecorvo:1957qd, MNS,
  Pontecorvo:1967fh, Gribov:1968kq}, plays an identical role and has
similar mathematical properties as the CKM matrix for quarks. The only
difference is the number of parameters that appear for the Majorana
neutrinos.

The least understood part of the SM is the Yukawa sector. In the
\textit{classical} SM (with masless neutrinos) it contains 13
phenomenological parameters: 9 quark and lepton masses and 4
parameters of the CKM matrix. Better understanding of the Yukawa
sector may lead to a significant reduction of the number of parameters
of the SM.

\section{General remarks about the  CKM and PMNS matrices}
\label{sec:gener-remarks-about}

The three most important results of our analysis are: 1.~The
CP~violation appears naturally and the CP~violating phase is not a
free parameter. 2.~The CKM and PMNS matrices have a common algebraic
structure and this extends the quark lepton universality. 3.~The
experimental data for the CKM and PMNS matrices suggest the $Z_{3}$
symmetry to be present in the Yukawa sector of the Standard Model.

The CKM matrix~$V$, when derived from the Yukawa couplings of quarks
with the Higgs field is equal to
\begin{equation}
  \label{eq:1}
  V=V_{L}^{u}{V_{L}^{d}}^{\dagger},
\end{equation}
where $V_{L}^{u}$ and $V_{L}^{d}$ are the unitary matrices that are
determined by the biunitary diagonalization of the Yukawa interactions
matrices of the \textit{up} and \textit{down} quarks.

It is not simple to obtain properties of the Yukawa couplings from the
experimental values of the CKM matrix~\citep{Workman:2022ynf}, because
there is an excessive arbitrariness in the splitting of the matrix $V$
in Eq.~\eqref{eq:1} into two unitary matrices. The conventional way to
derive those properties is to assume a discrete flavor symmetry or
some other properties, like textures for the Yukawa couplings and then
compare the predictions with experimental values for the CKM or PMNS
matrix elements. For a recent illustration of the first method see
Ref.~\citep{hernández2022predictive} and for the recent paper on
textures see~\citep{belfatto2023minimally}.

In phenomenological applications there are two mostly used
parameterizations of the CKM matrix: the standard
one~\citep{PhysRevLett.53.1802}, which is a superposition of three
rotations around the $x$, $y$ and $z$ axes with a complex rotation
around the~$y$ axis and a second one is the Wolfenstein
parameterization~\citep{PhysRevLett.51.1945}, which highlights the
hierarchy between the CKM matrix elements.

\section{New representation of the CKM and PMNS matrices}
\label{sec:new-repr-skm}

In this paper we will use a special representation of a general
unitary matrix as a product of three matrices
\begin{equation}
  \label{eq:2}
  U=O_{1}DO_{2}^{T}.
\end{equation}
Here $O_{1}$ and $O_{2}$ are orthogonal (rotation)
matrices~\footnote{For the 3-dimensional matrices we parameterize them
  by three angles, which are: two angles $\theta$ and $\phi$ which
  define the direction of the axis of rotation and the rotation angle
  $\alpha$ around the given axis.}  and~$D$ is the diagonal matrix
whose elements are imaginary exponents. A sketch of a proof that a
general unitary matrix can be expressed in a form~\eqref{eq:2} is
given in Ref.~\citep{decomposition}.

Now using the representation in Eq.~\eqref{eq:2} we will analyze the
CKM and PMNS (Dirac case) matrices~\footnote{The Majorana case will be
  discussed elsewhere.}. Our aim is not to find a new parameterization
of those matrices, but to show that the matrices $O_{1}$, $O_{2}$ and
$D$ have a special form, which can be identified with the symmetry present
in the Yukawa interactions of quarks and leptons. For this purpose we
will determine the matrices $O_{1}$, $O_{2}$ and $D$ from the
experimental input for the CKM and PMNS matrices using the method
described below.

The CKM and PMNS matrices have dimension $3\times 3$ and are
unitary. The rephasing freedom of the quark and lepton fields results
in the reduction to~4~of the number of significant parameters of each
matrix. A general $3\times 3$ unitary matrix has~9 free
parameters. This means that we have some freedom in the choice of the
values of the parameters in representation~\eqref{eq:2} for the
reproduction of the experimental values of the CKM and PMNS
matrices. In order to uncover the $Z_{3}$ symmetry of the Yukawa
sector we will make the following assumptions about the structure of
the matrices $O_{1}$, $O_{2}$ and $D$ in Eq.~(\ref{eq:2})
\begin{enumerate}
\item The diagonal matrix $D$ for \textit{both} CKM and PMNS matrices
  comprises of the third order roots of~1\linebreak ($Z_{3}$ symmetry) and is
  equal
  \begin{equation}
    \label{eq:3}
    D_{Z_{3}}=\operatorname{diag}(e^{\frac{4\pi i}{3}},e^{\frac{2\pi i}{3}},1).
  \end{equation}\discretionary{ma-}{trix}{matrix}
  The choice of such a form is also motivated by the successful
  two-angle parameterization of the CKM
  matrix~\citep{PhysRevLett.63.2189}, which uses a similar form of this
  matrix. This ansatz for the matrix~$D$ reduces the number of free
  parameters for each matrix by~3.
\item We choose the angles $\alpha$ of the rotation of the real
  orthogonal matrices $O_{1}$ and $O_{2}$ to be \textit{fixed} and
  equal to each other, so these matrices differ only by the axis of
  rotation and thus we remain with~4 free parameters for each matrix
  CKM and PMNS.
\end{enumerate}
These assumptions lead the following representation of the\linebreak
CKM and PMNS matrices
\begin{equation}
  \label{eq:4}
  V=\mathcal{R}(\alpha;\theta_{u},\phi_{u})D_{Z_{3}}\mathcal{R} (\alpha;\theta_{d},\phi_{d})^{T}.
\end{equation}
Here $\mathcal{R}(\alpha;\theta,\phi)$ is the matrix of rotation,
$\alpha$ is the angle of rotation and $\theta$ and $\phi$ are the
spherical angles of the axis of rotation.

\section{Parameters of the CKM and PMNS matrices from experiment}
\label{sec:determ-param-ckm}

Our goal is to investigate the compatibility of the
representation~\eqref{eq:4} with the measured values of the elements
of the CKM and PMNS matrices, by determination of the parameters of
the CKM matrix in representation~\eqref{eq:4} from the absolute values
of the matrix elements of the CKM and PMNS matrices generated by the
fit of these matrices to the experimental values. In such a way the
parameters of the unitary matrix in Eq.~\eqref{eq:4} will be
determined exactly.  As our input we use the central values of the
parameters of each matrix given by PDG~\citep{Workman:2022ynf}, which
are
\begin{equation}
  \label{eq:5}
  \begin{aligned}
    &\text{for the CKM matrix}\\
    &\lambda=0.22500,\; A=0.826,\; \bar{\rho}=0.159,\; \bar{\eta}=0.348, \\
    &\text{for the PMNS matrix (normal order)}\\
    &\sin^{2}(\theta_{12})=0.307,\;\sin^{2}(\theta_{23})=0.547,\\
    &\sin^{2}(\theta_{13})=0.0220, \;\delta_{\text{CP}}=1.23\pi\,\text{radians}\\
    &\text{for the PMNS matrix (inverted order)}\\
    &\sin^{2}(\theta_{12})=0.307,\;\sin^{2}(\theta_{23})=0.534,\\
    &\sin^{2}(\theta_{13})=0.0220, \;\delta_{\text{CP}}=1.23\pi\,\text{radians}
  \end{aligned}
\end{equation}
and calculate the absolute values of the elements of the CKM and PMNS
matrices. Next we determine \textit{exactly} the spherical angles
$\theta$ and $\phi$ for both matrices by solving the set of
corresponding equations for these absolute values.

As the next step in the search of the $Z_{3}$ symmetry we test the
assumption that the angle of rotation $\alpha$ is equal
\textit{exactly} to $2*\pi/3$.  Such an assumption is \textit{not
  compatible} with the experimental information for the CKM and PMNS
matrices. Subsequently we tested a small deviation of angle $\alpha$
from $2\pi/3$ for both matrices. The positive result of our search is
shown in Table~\ref{tab:1} where we give the values of the angle
$\alpha$ and spherical angles $\theta$ and $\phi$ which
\textit{exactly} reproduce the CKM and PMNS matrices generated from
the data given in Eq.~\eqref{eq:5}
\begin{table}[h]
  \centering
  \scalebox{0.93}{\begin{tabular}{l|c|c|c|c|c}
    Matrix&$\alpha$&$\theta_{u}$&$\phi_{u}$&$\theta_{d}$&$\phi_{d}$\\
    \hline
    CKM&$\frac{2\pi}{3}*1.04$&$52.92^{\circ}$&$46.35^{\circ}$ &$52.70^{\circ}$&$50.25^{\circ}$\\[5pt]
    PMNS (A)&$\frac{2\pi}{3}*1.01$&$64.24^{\circ}$&$138.66^{\circ}$&$156.96^{\circ}$ &$33.02^{\circ}$\\[5pt]
    PMNS (B)&$\frac{2\pi}{3}*1.01$&$64.18^{\circ}$&$138.98^{\circ}$&$156.55^{\circ}$ &$32.52^{\circ}$
  \end{tabular}}
  \caption{The angles $\alpha$, $\theta$ and $\phi$ of the CKM and
    PMNS matrices. PMNS~(A) stands for normal order and PMNS~(B)
    stands for the inverted order. Angle $\alpha$ is in radians and
    the remaining angles are in degrees.}
  \label{tab:1}
\end{table}
and we see that the deviation from the value $2\pi/3$ is small: for
the CKM matrix it is 4\% and for the PMNS matrix it is only 1\%. The
striking fact is that none of the angles in Table~\ref{tab:1} are
small, which suggests a non perturbative nature of the CKM and PMNS
matrices.

\section{Discussion of the results}
\label{sec:discussion-results}

The data in Table~\ref{tab:1} show that the spherical angles of the
axes of rotation in the CKM matrix for the \textit{up} and
\textit{down} quarks are almost equal and this is a consequence of the
fact that the matrix of the absolute values CKM matrix is
\textit{almost} symmetric. The spherical angles of the axes of
rotation of the CKM and PMNS matrices are different, but this was to
be expected, because these matrices are not equal.

The structure of the matrix $D_{Z_{3}}$ and the values of the
angle~$\alpha$ for the CKM and PMNS matrices give a strong support for
the idea of the presence of the $Z_{3}$ symmetry for the matrices CKM
and PMNS and thus also for the Yukawa couplings of quarks and
leptons. A deviation of the angle $\alpha$ from the value $2*\pi/3$
may be caused by a small $Z_{3}$ symmetry breaking.

Our representation of the CKM and PMNS matrices allows the
determination of the \textit{left} bi-unitary diagonalizing matrices
$V_{L}^{u}$ and $V_{L}^{d}$
\begin{equation}
  \label{eq:7}
  V_{L}^{f}=\mathcal{R}(\alpha,\theta_{f},\phi_{f})D_{f},\quad
  f=u,d,
\end{equation}
where $D_{u}$ and $D_{d}$ are the diagonal matrices
\begin{equation}
  \label{eq:8}
  D_{u}=\operatorname{diag}(1,e^{2\pi i/3},e^{4\pi i/3}),\;\;
  D_{d}=\operatorname{diag}(e^{2\pi i/3},1,e^{4\pi i/3})
\end{equation}
and
\begin{equation}
  \label{eq:9}
  D_{Z_{3}}=D_{u}D_{d}^{\dagger}.
\end{equation}
It follows from the biunitary diagonalization that the quark Yu\-ka\-wa
couplings matrices $Y^{u}$ and $Y^{d}$ are equal
\begin{multline}
  \label{eq:10}
  \frac{v}{\sqrt{2}}Y^{f}={V_{L}^{f}}^{\dagger}M^{f}V_{R}^{f}\\
  = D_{f}^{\dagger}\mathcal{R}(\alpha,\theta_{f},\phi_{f})^{T} M^{f}V_{R}^{f},
  \quad f=u,d.
\end{multline}
Here $M^{f}$ is the diagonal quark or lepton mass matrix and $v$ is
the vacuum expectation value of the Higgs field. The \textit{right}
bi-unitary diagonalizing matrices do not enter into any observables and we will consider two cases: $V_{R}^{f}=I$ or $V_{R}^{f}=V_{L}^{f}$. These choices lead to different results for the matrices of the quark and lepton Yukawa couplings:
\begin{equation}
  \label{eq:16} \frac{v}{\sqrt{2}}Y^{f}=
  \begin{cases}
 \text{case A} &D_{f}^{\dagger}\mathcal{R}(\alpha,\theta_{f},\phi_{f})^{T} M^{f} \text{ for }V_{R}^{u}=I\\
\text{case B}  & D_{f}^{\dagger}\mathcal{R}(\alpha,\theta_{f},\phi_{f})^{T} M^{f}V_{L}^{f}D_{f}   \text{ for }V_{R}^{f}=V_{L}^{f}
  \end{cases}
\end{equation}
% \begin{equation}
%   \label{eq:16} \frac{v}{\sqrt{2}}Y^{f}=
%   \begin{cases}
%  D_{f}^{\dagger}\mathcal{R}(\alpha,\theta_{f},\phi_{f})^{T} M^{f}&  \text{casa A for }V_{R}^{u}=I\\
%      D_{f}^{\dagger}\mathcal{R}(\alpha,\theta_{f},\phi_{f})^{T} M^{f}V_{L}^{f}D_{f} &  \text{case B for }V_{R}^{f}=V_{L}^{f}
%   \end{cases}
% \end{equation}

Below we give the explicit numerical values of the $V_{l}^{f}$ and
$Y^{f}$ matrices for quarks obtained from the values of the angles in
Table~\ref{tab:1}. For clarity we did not substitute the phases
$e^{2\pi i/3}$ and $e^{4\pi i/3}$ by its numerical values, because in
such a way the mathematical structure of these matrices becomes clear.
\begin{gather}
    \label{eq:14}
    V_{L}^{u}=
    \begin{pmatrix}
      -0.094 & 0.0043\cdot e^{2\pi i/3} & 1.00\cdot e^{4\pi i/3} \\
      0.99 & -0.047\cdot e^{2\pi i/3} & 0.094\cdot e^{4\pi i/3} \\
      0.048 & 1.00\cdot e^{2\pi i/3} & 0.00021\cdot e^{4\pi i/3} \\
    \end{pmatrix}\\[6pt]
  \label{eq:15}
  V_{L}^{d}=  
\begin{pmatrix}
 -0.16\cdot e^{2\pi i/3} & -0.0088 & 0.99\cdot e^{4\pi i/3} \\
 0.99\cdot e^{2\pi i/3} & 0.017 & 0.16\cdot e^{4\pi i/3} \\
 -0.018\cdot e^{2\pi i/3} & 1.00 & 0.0069\cdot e^{4\pi i/3} \\
\end{pmatrix}
\end{gather}
These matrices have the following properties
\begin{itemize}
\item The numerical structure of the matrices $V_{L}^{u}$ and $V_{L}^{d}$ is similar.
\item The phases of the matrix elements in each column are equal.
\item There is a hierarchy between the matrix elements of the matrices: the matrix elements: in each row (or column) one matrix element is of the order $\sim1$ and the remaining matrix elements are small.
\item The hierarchy of the matrix elements is \textit{independent} of the masses of quarks.
\item There are no exact textures.
\end{itemize}

Next, we calculate from Eq.~\eqref{eq:10} the matrices of the Yukawa
couplings $Y^{f}$ for two cases explained in Eq.~\eqref{eq:16}. In
the equations below we provide the numerical values of the matrices
$vY^{u,d}/(m_{t,b}\sqrt{2})$ for the \textit{up} and \textit{down}
quarks. For $Y^{u}$ and $Y^{d}$ we normalize the values of these
matrices by the masses $m_{t}$ and $m_{b}$, respectively.
\begin{subequations}
  \begin{multline}
    \label{eq:6}
    \text{case A: }\quad
    \frac{v}{m_{t}\sqrt{2}}Y^{u}\\ =
    \begin{pmatrix}
      -1.18\cdot 10^{-6} & 0.0073 & 0.048 \\
      5.35\cdot 10^{-8} \cdot e^{4\pi i/3} & -0.00035 \cdot e^{4\pi i/3} &
                                                                           1.00 \cdot e^{4\pi i/3} \\
      0.000012 \cdot e^{2\pi i/3} & 0.00069 \cdot e^{2\pi i/3} & 0.00021 \cdot e^{2\pi i/3}
    \end{pmatrix}
  \end{multline}
  \begin{multline}
    \label{eq:12}
    \text{case B: }\quad  
    \frac{v}{m_{t}\sqrt{2}}Y^{u}\\ =  
    \begin{pmatrix}
      0.0095 & 0.047 \cdot e^{2\pi i/3} & 0.00070 \cdot e^{4\pi i/3} \\
      0.047 \cdot e^{4\pi i/3} & 1.00  & 0.00018 \cdot e^{2\pi i/3} \\
      0.00070 \cdot e^{2\pi i/3} & 0.00018 \cdot e^{4\pi i/3} & 0.000078
    \end{pmatrix}
  \end{multline}
\end{subequations}

\begin{subequations}
  \begin{multline}
    \label{eq:11}
    \text{case A: }\quad  
    \frac{v}{m_{b}\sqrt{2}}Y^{d}\\ =  
    \begin{pmatrix}
      -0.00018 \cdot e^{4\pi i/3} & 0.022 \cdot e^{4\pi i/3} & -0.018 \cdot e^{4\pi i/3} \\
      -9.87\cdot 10^{-6} & 0.00038 & 1.00 \\
      0.0011 \cdot e^{2\pi i/3} & 0.0037 \cdot e^{2\pi i/3} & 0.0060 \cdot e^{2\pi i/3}
    \end{pmatrix}
  \end{multline}
  \begin{multline}
    \label{eq:13}
     \text{case B: }\quad    
    \frac{v}{m_{b}\sqrt{2}}Y^{d}\\ =  
    \begin{pmatrix}
      0.022  & -0.018 \cdot e^{4\pi i/3} & 0.0033 \cdot e^{2\pi i/3} \\
      -0.018 \cdot e^{2\pi i/3} & 1.00 & 0.0060 \cdot e^{4\pi i/3} \\
      0.0033 \cdot e^{4\pi i/3} & 0.0060 \cdot e^{2\pi i/3} & 0.0017  
    \end{pmatrix}
  \end{multline}
\end{subequations}
We observe that the Yukawa couplings $Y^{f}$ have the following
properties
\begin{itemize}
\item For case~A the phases of the matrix elements in each row are
  equal and for case B the matrices are hermitian.
\item There are no \textit{exact} textures, because all matrix
  elements of $\mathcal{R}(\alpha,\theta_{f},\phi_{f})$ are
  non-vanishing.
\item The absolute value of only one matrix element of each matrix is
  close to~1, the remaining ones are several order of magnitude
  smaller. For case~A, the matrix element $(2,3)$ dominates, and for
  case~B, the element $(2,2)$ dominates.
\item The hierarchy in the $Y^{f}$ matrices is stronger than in case
  of the $V_{L}^{f}$ matrices which is caused by the strong hierarchy
  of the quark masses.
\end{itemize}

\section{Conclusions}
\label{sec:conclusions}

Let us summarize the obtained results by stating that we formulated a
new representation of the CKM and PMNS matrices with the following
properties
\begin{enumerate}
\item \textit{Both} matrices CKM and PMNS have the following identical
  mathematical structure
  \begin{equation*}
    V_{\text{CKM,PMNS}}=\mathcal{R}(\alpha,\theta_{u},\phi_{u})D_{Z_{3}}
    \mathcal{R}(\alpha,\theta_{d},\phi_{d})^{T}
  \end{equation*}
  with $D_{Z_{3}}$ fixed, given in Eq.~\eqref{eq:3} and the angle
  $\alpha$ is close to $2\pi/3$.
\item The structure of the diagonal matrix $D_{Z_{3}}$, whose elements
  are cubic roots of~1 and the value of the angle $\alpha$ close to
  $2\pi/3$ support the idea of the presence of the $Z_{3}$ symmetry in
  the Yukawa sector of the Standard Model.
\item The CP violation phase in the CKM matrix \textit{is not} a
  fitted parameter, but is the consequence of the structure of the
  phases of the Yukawa couplings matrices (see Eq.~\ref{eq:10}). This
  may be a starting point of a future theory of CP violation.
\item The angles describing the diagonalizing matrices $V_{L}^{u}$ and
  $V_{L}^{d}$ are not small (see Table~\ref{tab:1}). This means that
  the hierarchy present in the CKM matrix seems to be accidental and
  has a different origin than the hierarchy of the quark and lepton
  masses. This point of view is supported by the lack of the hierarchy
  in the PMNS matrix.
\item The Yukawa couplings do not have \textit{exact} textures.
\end{enumerate}
Points 1., 2.\ and 3.\ provide a description of CP violation without
free parameters in the SM and reestablish the quark-lepton
universality, which seemed missing in the Yukawa
sector. Points~4.\ and~5.\ may be a good testing ground in search of the
quark flavor symmetry.

\section*{Acknowledgements}
This work is supported in part by Proyecto SIP20231470, Secretaría de
Investigación y Posgrado, Beca EDI y Comisión de Operación y Fomento
de Actividades Académicas (COFAA) del Instituto Politécnico Nacional
(IPN), Mexico. We also thank professor Xing Zhi Zhong for constructive
comments.

% \bibliographystyle{elsarticle-num} 
% \bibliography{cp}

\begin{thebibliography}{10}
\expandafter\ifx\csname url\endcsname\relax
  \def\url#1{\texttt{#1}}\fi
\expandafter\ifx\csname urlprefix\endcsname\relax\def\urlprefix{URL }\fi
\expandafter\ifx\csname href\endcsname\relax
  \def\href#1#2{#2} \def\path#1{#1}\fi

\bibitem{PhysRev.105.1413}
C.~S. Wu, E.~Ambler, R.~W. Hayward, D.~D. Hoppes, R.~P. Hudson,
  \href{https://link.aps.org/doi/10.1103/PhysRev.105.1413}{Experimental {T}est
  of {P}arity {C}onservation in {B}eta {D}ecay}, Phys. Rev. 105 (1957)
  1413--1415.
\newblock \href {https://doi.org/10.1103/PhysRev.105.1413}
  {\path{doi:10.1103/PhysRev.105.1413}}.
\newline\urlprefix\url{https://link.aps.org/doi/10.1103/PhysRev.105.1413}

\bibitem{PhysRev.104.254}
T.~D. Lee, C.~N. Yang,
  \href{https://link.aps.org/doi/10.1103/PhysRev.104.254}{Question of {P}arity
  {C}onservation in {W}eak {I}nteractions}, Phys. Rev. 104 (1956) 254--258.
\newblock \href {https://doi.org/10.1103/PhysRev.104.254}
  {\path{doi:10.1103/PhysRev.104.254}}.
\newline\urlprefix\url{https://link.aps.org/doi/10.1103/PhysRev.104.254}

\bibitem{PhysRev.105.1415}
R.~L. Garwin, L.~M. Lederman, M.~Weinrich,
  \href{https://link.aps.org/doi/10.1103/PhysRev.105.1415}{Observations of the
  {F}ailure of {C}onservation of {P}arity and {C}harge {C}onjugation in {M}eson
  {D}ecays: the {M}agnetic {M}oment of the {F}ree {M}uon}, Phys. Rev. 105
  (1957) 1415--1417.
\newblock \href {https://doi.org/10.1103/PhysRev.105.1415}
  {\path{doi:10.1103/PhysRev.105.1415}}.
\newline\urlprefix\url{https://link.aps.org/doi/10.1103/PhysRev.105.1415}

\bibitem{PhysRevLett.13.138}
J.~H. Christenson, J.~W. Cronin, V.~L. Fitch, R.~Turlay,
  \href{https://link.aps.org/doi/10.1103/PhysRevLett.13.138}{Evidence for the
  $2\ensuremath{\pi}$ {D}ecay of the ${K}_{2}^{0}$ {M}eson}, Phys. Rev. Lett.
  13 (1964) 138--140.
\newblock \href {https://doi.org/10.1103/PhysRevLett.13.138}
  {\path{doi:10.1103/PhysRevLett.13.138}}.
\newline\urlprefix\url{https://link.aps.org/doi/10.1103/PhysRevLett.13.138}

\bibitem{199843}
{A. Angelopoulos et al.},
  \href{https://www.sciencedirect.com/science/article/pii/S0370269398013562}{First
  direct observation of time-reversal non-invariance in the neutral-kaon
  system}, Physics Letters B 444~(1) (1998) 43--51.
\newblock \href {https://doi.org/https://doi.org/10.1016/S0370-2693(98)01356-2}
  {\path{doi:https://doi.org/10.1016/S0370-2693(98)01356-2}}.
\newline\urlprefix\url{https://www.sciencedirect.com/science/article/pii/S0370269398013562}

\bibitem{Fermi1934}
E.~Fermi, \href{https://doi.org/10.1007/BF01351864}{{Versuch einer Theorie der
  $\beta$-Strahlen. I}}, Zeitschrift f{\"u}r Physik 88~(3) (1934) 161--177.
\newblock \href {https://doi.org/10.1007/BF01351864}
  {\path{doi:10.1007/BF01351864}}.
\newline\urlprefix\url{https://doi.org/10.1007/BF01351864}

\bibitem{10.1119/1.1974382}
F.~L. Wilson, \href{https://doi.org/10.1119/1.1974382}{{Fermi's Theory of Beta
  Decay}}, American Journal of Physics 36~(12) (1968) 1150--1160, a complete
  English translation is given of the classic Enrico Fermi paper on beta decay
  published in Zeitschrift für Physik in 1934.
\newblock \href
  {http://arxiv.org/abs/https://pubs.aip.org/aapt/ajp/article-pdf/36/12/1150/11891052/1150\_1\_online.pdf}
  {\path{arXiv:https://pubs.aip.org/aapt/ajp/article-pdf/36/12/1150/11891052/1150\_1\_online.pdf}},
  \href {https://doi.org/10.1119/1.1974382} {\path{doi:10.1119/1.1974382}}.
\newline\urlprefix\url{https://doi.org/10.1119/1.1974382}

\bibitem{Sudarshan}
E.~C.~G. Sudarshan, R.~E. Marshak, {The Nature of the Four Fermion
  Interaction}, in: N.~Zanichelli (Ed.), Proc. of the Padua-Venice Conference
  on Mesons and Newly-Discovered Particles, September 1957, Societ\`a Italiana
  di Fisica, 1958, pp. 508--515.

\bibitem{PhysRev.109.1860.2}
E.~C.~G. Sudarshan, R.~E. Marshak,
  \href{https://link.aps.org/doi/10.1103/PhysRev.109.1860.2}{{Chirality
  Invariance and the Universal Fermi Interaction}}, Phys. Rev. 109 (1958)
  1860--1862.
\newblock \href {https://doi.org/10.1103/PhysRev.109.1860.2}
  {\path{doi:10.1103/PhysRev.109.1860.2}}.
\newline\urlprefix\url{https://link.aps.org/doi/10.1103/PhysRev.109.1860.2}

\bibitem{PhysRev.109.193}
R.~P. Feynman, M.~Gell-Mann,
  \href{https://link.aps.org/doi/10.1103/PhysRev.109.193}{{Theory of the Fermi
  Interaction}}, Phys. Rev. 109 (1958) 193--198.
\newblock \href {https://doi.org/10.1103/PhysRev.109.193}
  {\path{doi:10.1103/PhysRev.109.193}}.
\newline\urlprefix\url{https://link.aps.org/doi/10.1103/PhysRev.109.193}

\bibitem{PhysRevLett.10.531}
N.~Cabibbo, \href{https://link.aps.org/doi/10.1103/PhysRevLett.10.531}{{Unitary
  Symmetry and Leptonic Decays}}, Phys. Rev. Lett. 10 (1963) 531--533.
\newblock \href {https://doi.org/10.1103/PhysRevLett.10.531}
  {\path{doi:10.1103/PhysRevLett.10.531}}.
\newline\urlprefix\url{https://link.aps.org/doi/10.1103/PhysRevLett.10.531}

\bibitem{10.1143/PTP.49.652}
M.~Kobayashi, T.~Maskawa,
  \href{https://doi.org/10.1143/PTP.49.652}{{CP-Violation in the Renormalizable
  Theory of Weak Interaction}}, Progress of Theoretical Physics 49~(2) (1973)
  652--657.
\newblock \href
  {http://arxiv.org/abs/https://academic.oup.com/ptp/article-pdf/49/2/652/5257692/49-2-652.pdf}
  {\path{arXiv:https://academic.oup.com/ptp/article-pdf/49/2/652/5257692/49-2-652.pdf}},
  \href {https://doi.org/10.1143/PTP.49.652} {\path{doi:10.1143/PTP.49.652}}.
\newline\urlprefix\url{https://doi.org/10.1143/PTP.49.652}

\bibitem{Pontecorvo:1957cp}
B.~Pontecorvo, {Mesonium and anti-mesonium}, Sov. Phys. JETP 6 (1957) 429,
  [Zh.Eksp.Teor.Fiz. 33 (1957) 549-551].

\bibitem{Pontecorvo:1957qd}
B.~Pontecorvo, {Inverse beta processes and nonconservation of lepton charge},
  Sov. Phys. JETP 7 (1958) 172--173, [Zh.Eksp.Teor.Fiz. 34 (1957) 247].

\bibitem{MNS}
Z.~Maki, M.~Nakagawa, S.~Sakata, {Remarks on the Unified Model of Elementary
  Particles}, Progress of Theoretical Physics 28 (1962) 870--880.

\bibitem{Pontecorvo:1967fh}
B.~Pontecorvo, {Neutrino Experiments and the Problem of Conservation of
  Leptonic Charge}, Sov. Phys. JETP 26 (1968) 984--988, [Zh.Eksp.Teor.Fiz. 53
  (1967) 1717--1725].

\bibitem{Gribov:1968kq}
V.~N. Gribov, B.~Pontecorvo, {Neutrino astronomy and lepton charge}, Phys.
  Lett. B 28 (1969) 493--496.
\newblock \href {https://doi.org/10.1016/0370-2693(69)90525-5}
  {\path{doi:10.1016/0370-2693(69)90525-5}}.

\bibitem{Workman:2022ynf}
R.~L. Workman, Others, {Review of Particle Physics}, PTEP 2022 (2022) 083C01.
\newblock \href {https://doi.org/10.1093/ptep/ptac097}
  {\path{doi:10.1093/ptep/ptac097}}.

\bibitem{hernández2022predictive}
A.~E.~C. Hernández, C.~Espinoza, J.~C. Gómez-Izquierdo, J.~M. González,
  M.~Mondragón, {Predictive extended 3HDM with $S_4$ family symmetry} (2022).
\newblock \href {http://arxiv.org/abs/2212.12000} {\path{arXiv:2212.12000}}.

\bibitem{belfatto2023minimally}
B.~Belfatto, Z.~Berezhiani, {Minimally modified Fritzsch texture for quark
  masses and CKM mixing} (2023).
\newblock \href {http://arxiv.org/abs/2305.00069} {\path{arXiv:2305.00069}}.

\bibitem{PhysRevLett.53.1802}
L.-L. Chau, W.-Y. Keung,
  \href{https://link.aps.org/doi/10.1103/PhysRevLett.53.1802}{Comments on the
  parametrization of the kobayashi-maskawa matrix}, Phys. Rev. Lett. 53 (1984)
  1802--1805.
\newblock \href {https://doi.org/10.1103/PhysRevLett.53.1802}
  {\path{doi:10.1103/PhysRevLett.53.1802}}.
\newline\urlprefix\url{https://link.aps.org/doi/10.1103/PhysRevLett.53.1802}

\bibitem{PhysRevLett.51.1945}
L.~Wolfenstein,
  \href{https://link.aps.org/doi/10.1103/PhysRevLett.51.1945}{Parametrization
  of the kobayashi-maskawa matrix}, Phys. Rev. Lett. 51 (1983) 1945--1947.
\newblock \href {https://doi.org/10.1103/PhysRevLett.51.1945}
  {\path{doi:10.1103/PhysRevLett.51.1945}}.
\newline\urlprefix\url{https://link.aps.org/doi/10.1103/PhysRevLett.51.1945}

\bibitem{decomposition}
{Mathematics Stack Exchange: orangeskid}, Unitary matrix decomposition using
  orthogonal matrices,
  https://math.stackexchange.com/questions/1127779/unitary-matrix-decomposition-using-orthogonal-matrices,
  {A}ccessed: 2023-08-05 (2015).

\bibitem{PhysRevLett.63.2189}
P.~Kielanowski,
  \href{https://link.aps.org/doi/10.1103/PhysRevLett.63.2189}{{Two-angle
  parametrization of the Kobayashi-Maskawa matrix: A relation between CP
  violation and the Cabibbo-type angles}}, Phys. Rev. Lett. 63 (1989)
  2189--2191.
\newblock \href {https://doi.org/10.1103/PhysRevLett.63.2189}
  {\path{doi:10.1103/PhysRevLett.63.2189}}.
\newline\urlprefix\url{https://link.aps.org/doi/10.1103/PhysRevLett.63.2189}

\end{thebibliography}

\end{document}